\documentclass[twocolumn,superscriptaddress,showpacs,aps,prb]{revtex4}
\usepackage{makeidx}
\usepackage{bm}
\usepackage{graphics}
\usepackage[dvips]{epsfig}
\newcounter{fig}

\begin{document}
\title{Quantum magneto-optics  of graphite with trigonal warping }
\author{L.A. Falkovsky}
\affiliation{L.D. Landau Institute for Theoretical Physics, Moscow
119334, Russia} \affiliation{Institute of the High Pressure
Physics, Troitsk 142190, Russia}
\pacs{76.40.+b, 78.20.-e, 81.05.uf}
\date{\today}      

\begin{abstract}
The optical conductivity of graphite in quantizing magnetic fields is studied.
Both the dynamical  conductivities, longitudinal as well as  Hall's, are 
analytically evaluated. The conductivity peaks are explained in terms of electron transitions. We have shown that the trigonal warping in graphite can be considered within the perturbation theory at the strong magnetic field larger than 1 T approximately. The main optical transitions obey the selection rule with $\Delta n=1$ for the Landau number $n$, however the $\Delta n=2$ transitions due to the trigonal warping with the small probability are also essential. 
The Kerr rotation and reflectivity in graphite in the quantizing magnetic fields are calculated. 
Parameters of the Slonczewski--Weiss--McClure   model  are used in the fit  taking into account the previous dHvA measurements  and correcting some of them for the case of the strong magnetic fields.
\end{abstract}
\maketitle
\section{introduction}
 Properties of graphite have attracted much attention for more than 50 years. Many of that properties were successfully explained within the Slonczewski--Weiss--McClure (SWMC) theory \cite{SW}. The most accurate method to study the band structure of graphite is a study of the Landau levels (LLs)  through experiments such as magneto-optics \cite{ST,TDD,DDS,MMD,LTP,OFM,OFS,OP,CLW} and magneto-transport  \cite{KTS,LK,JZS,SOP,RM}. However, the interpretation of the experimental results involves a significant degree of uncertainty since, as it is not clear how the resonances
 should be identified and which electron transitions they correspond to.
 
The SWMC theory requires
the use of many tight-binding parameters and provides the simple description of observed phenomena either in the semiclassical limit of week magnetic fields or for high frequencies when the largest tight-binding inter-layer parameter $\gamma_1$  plays the  leading role \cite{Fal}.  It is more difficult to take into account the parameter $\gamma_3$ known as "trigonal warping". 
   Usually, it is either neglected \cite{OP,Fal,LA,CBN,ZLB} or considered numerically \cite{UUU,Na,PP,GAW,KCD}. The Bohr-Sommerfeld quantization condition was applied in Ref. \cite{Dr} to find in low magnetic fields the level structure including the trigonal warping. In any case, only the problem of levels was considered so far, and no calculations of 
   conductivities were done in order to evaluate the optic properties of graphite. The problem appearing  for  three-dimensional systems in the magnetic field connects partly with integrating  over the momentum projection $k_z$ along the magnetic field.   
   
   The SWMC model can be simplified assuming that only the integration limits such as the $K$ and $H$ points in the Brillouin zone  produce the main contributions \cite{OFS,LA,CBN}. 
   Such an approximation is similar to the theory of magneto-optical effects in topological insulators \cite{TM} and graphene \cite{MHA}.   However,
 the band extrema or the integration limits at the Fermi level  can also contribute into the absorption. Therefore,
the analytical expression for the dynamic conductivity in
the presence of magnetic fields is needed for the interpretation of the magneto-optics experiments. 

In this paper, motivated by the experimental study of the Faraday rotation in single- and multilayer graphene \cite{CLW}, we propose a theory of magneto-optics phenomena in graphite in strong magnetic fields including the interlayer hopping parameters $\gamma_3$ and $\gamma_4$ in the Hamiltonian.  The trigonal warping $\gamma_3$ is considered as a perturbation with the help of the Green's function method.  Not only the energy-level structure corrected due to the trigonal warping  is found, but
 the expressions for the longitudinal and Hall dynamical conductivities are derived.  Our main theoretical finding is the 
reflectivity and the Kerr angle   for graphite  in strong magnetic fields.
\section{Landau levels in graphite with trigonal warping}
Taking into account  the tight-binding parameters of the SWMC theory,  
the effective Hamiltonian near the $ KH$ line of the Brillouin zone in graphite writes in the form of Ref. \cite{PP,GAW}
\begin{equation}
H(\mathbf{k})=\left(
\begin{array}{cccc}
\tilde{\gamma}_5     & vk_{+} & \tilde{\gamma}_1 &\tilde{\gamma}_4vk_{-}/\gamma_0\\
vk_{-} & \tilde{\gamma}_2  & \tilde{\gamma}_4vk_{-}/\gamma_0& \tilde{\gamma}_3vk_{+}/\gamma_0\\
\tilde{\gamma}_1   &\tilde{\gamma}_4vk_{+}/\gamma_0 & \tilde{\gamma}_5   &vk_{-}\\
\tilde{\gamma}_4vk_{+}/\gamma_0 & \tilde{\gamma}_3vk_{-}/\gamma_0 &vk_{+} &\tilde{\gamma}_2
\end{array}%
\right)\,,   \label{ham}
\end{equation}%
where $k_{\pm}=\mp ik_x-k_y$ are the momentum components and  $v$ is the velocity parameter  in the  intra-layer directions; 
$\tilde{\gamma}_j$ are the functions of the  $k_z$ momentum in the main axis direction
\begin{eqnarray}\tilde{\gamma_2}=2\gamma_2\cos{(2k_zd_0)}\,,\tilde{\gamma}_5=2\gamma_5\cos{(2k_zd_0)}+\Delta\,,\nonumber\\ 
\tilde{\gamma}_i=2\gamma_i\cos{(k_zd_0)}\quad\text{for}\quad i=1,3,4,\nonumber  
\end{eqnarray}
with the distance $d_0=3.35$ \AA\, between  the layers in graphite. The nearest-neighbor hopping integral $\gamma_0\approx 3$ eV corresponds with the in-layer inter-atomic distance $a_0=1.415 ~\AA$  and the Fermi velocity parameter $v=1.5a_0\gamma_0 = 10^6$ m/s.

For the zero magnetic field, the eigenvalues of the Hamiltonian give four close bands. 
In the magnetic field $B$, the momentum projections $k_{x,y}$ become the operators obeying  the commutation rule $\{\hat{k}_{+},\hat{k}_{-}\}=-2e\hbar B /c$, and we  use the relations
\[\hat{k}_+=\sqrt{2|e|\hbar B/c}\,a, \quad  \hat{k}_-=\sqrt{2|e|\hbar B/c}\,a^+\,,\]
involving the creation $a^+$ and annihilation operators $a$. We will write only one of two $x,y$ space coordinates including the corresponding degeneracy proportional to the magnetic field in the final results.

\begin{table}[]
\caption{\label{tb1} The parameters of the Hamiltonian, Eq. (\ref{ham}), their values in the SWMC model, and obtained in the experimental works, all in meV.  }
        \begin{ruledtabular}
                \begin{tabular}{cccccccccc}
 &(\ref{ham})&$\gamma_0$ & $\gamma_1$ & $\gamma_2$& $\gamma_3$ & $\gamma_4$ & $\gamma_5$ & $\Delta$ & $\varepsilon_F$\\
 & &3050& 360&$-10.2$&270&$-150$&$-1.5$&16&$-4.1$\\
 \hline
 &S$^a$& $\gamma_0$ & $\gamma_1$ & $2\gamma_2$& $\gamma_3$ & $-\gamma_4$ & $2\gamma_5$ & $\Delta+2(\gamma_2-\gamma_5)$&  2$\gamma_2+\varepsilon_F$\\  
&M$^b$& 3160& 390&$-20$&276&44&38&8&$-24$\\ 
&D$^c$& 3120& 380&$-21$&315&120&$-3$&$-2$&$-$\\ 
\end{tabular}
\end{ruledtabular}
 $^a$SWMC,Ref. \cite{SW},
 $^b$Mendez et al, Ref. \cite{MMD},
 $^c$Doezema et al, Ref. \cite{DDS}.
\end{table} 

We search the eigenfunction of the Hamiltonian (\ref{ham}) in the form
\begin{equation}
\psi_{sn}^{\alpha}(x)=
\left\{\begin{array}{c}
 C^{1}_{sn}\varphi_{n-1}(x)\\
 C^{2}_{sn}\varphi_{n}(x)\\
 C^{3}_{sn}\varphi_{n-1}(x)\\
 C^{4}_{sn}\varphi_{n-2}(x)\,
\end{array}\right.\,,\label{func}
\end{equation}
where  $\varphi_{n}(x)$ are 
orthonormal Hermitian   functions with the Landau numbers $n\ge0$.  One sees that every row
in the Hamiltonian (\ref{ham}) becomes proportional to the definite Hermitian function if the terms with $\gamma_3$ is omitting. We will show that the terms proportional to $\gamma_3/\gamma_0$ can  be   considered  within the perturbation theory at strong magnetic fields.
 \begin{figure}[]
\resizebox{.52\textwidth}{!}{\includegraphics{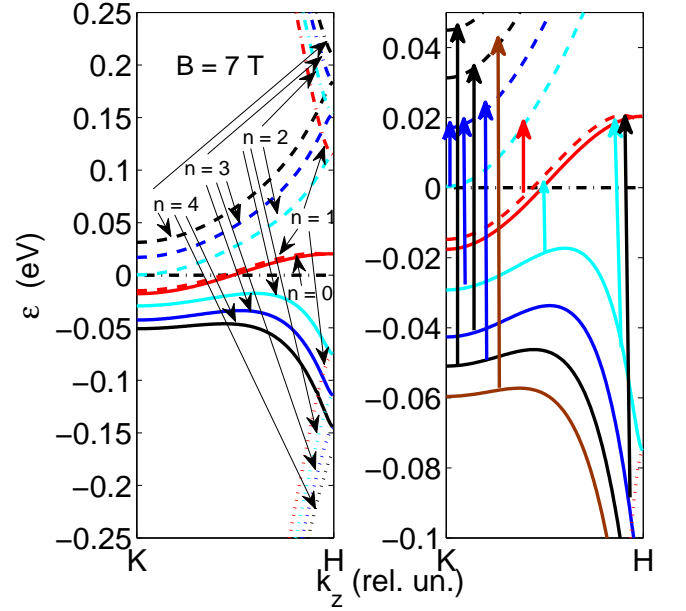}}
\caption{(Color online) LLs $\varepsilon_{sn}$ for $n$ = 0 to 4 in four bands $s=1,2,3,4$ (in dotted, solid, dashed, and dash-dotted lines, correspondingly) as functions of momentum $k_z$ along the $KH$ line in the Brillouin zone 
($K=0,\, H = \pi/2d_0$) at  the magnetic field $B$ = 7~T with the SWMC model parameters given in Tabl. 1.
 The main electron transitions shown in the right panel below 100 meV are possible between the levels with the selection rule  $\Delta n =\pm 1$\,, see text. }
\label{d7strf}\end{figure}

Canceling  the Hermitian functions from the equations,  we obtain 
a system of the linear equations for the eigenvector~${\bf C}_{sn}$
\begin{equation}
\left(
\begin{array}{cccc}
\tilde{\gamma}_5-\varepsilon     & \omega_c\sqrt{n} & \tilde{\gamma}_1  & 
\omega_4\sqrt{n-1}\\
\omega_c\sqrt{n} & \tilde{\gamma}_2-\varepsilon      & \omega_4\sqrt{n}& 0\\
\tilde{\gamma}_1      &\omega_4\sqrt{n} & \tilde{\gamma}_5-\varepsilon   &\omega_c\sqrt{n-1}\\
\omega_4\sqrt{n-1}& 0 &\omega_c\sqrt{n-1} &\tilde{\gamma}_2-\varepsilon
\end{array}%
\right) \times\left\{\begin{array}{c}
C^{1}_{sn}\\
C^{2}_{sn}\\
C^{3}_{sn}\\
C^{4}_{sn}
\end{array} \right.=0\,  \label{ham2}
\end{equation}
where the band number $s=1,2,3,4$ numerates the solutions at given $n$  from the bottom, $\omega_c=v\sqrt{2|e|\hbar B/c}$\,  and\, $\omega_{4}=\tilde{\gamma}_{4}\omega_c/\gamma_0$.
 
The eigenvalues of the matrix in Eq. (\ref{ham2}) are easily found,  they are shown in Fig. \ref{d7strf} as a function of the momentum $k_z$. For each Landau number  $n\ge 2$ and momentum $k_z$, there are four eigenvalues
 $\varepsilon_{s}(n)$  and four corresponding eigenvectors, Eq.  (\ref{func}), marked by  the band subscript  $s$.    We will use the notation $|sn\rangle$ for levels.
  In addition, there are four  levels. One of them, \begin{equation}\varepsilon_1(n=0)=\tilde{\gamma}_2\label{n0}\end{equation}
 for $n=0$ with the eigenvector ${\bf C}_0=(0,1,0,0)$ as is evident from Eq. (\ref{func}). It  intersects   the Fermi level
 and belongs to the  electron (hole) band near the $K$     $(H)$ point.  Others  three levels  indicated with $n=1$  and $s=1,2,3$ are determined by first three equations of the system (\ref{ham2}) with $C^4_{s1}=0$.
  The $|21\rangle$ level is  close to the $|10\rangle$ level. In the  region of $k_z$,  $\gamma_1/\cos{z}\gg \gamma_2$, where the electrons are located, this level has the energy 
\[\varepsilon_2(n=1)=\tilde{\gamma}_2-2\frac{\omega_c^2\tilde{\gamma}_4}{\tilde{\gamma}_1\gamma_0}\,.\]
 In the same region, the two closest bands ($s=2,3$) with 
 $n\ge 2$ are written as
 \begin {equation}
\begin{array}{c}
\varepsilon_{2,3}(n)={\displaystyle\tilde{\gamma}_2-\frac{\omega_c^2\tilde{\gamma}_4}{\tilde{\gamma}_1\gamma_0}(2n-1)} 
{\displaystyle\mp
\frac{\omega_c^2}{\tilde{\gamma}_1}\sqrt{n(n-1)}}\,
\end{array}\label{de1}\end{equation}
 within accuracy of $(\tilde{\gamma}_4/\gamma_0)^2$.
 
 \begin{figure}[]
\resizebox{.25\textwidth}{!}{\includegraphics{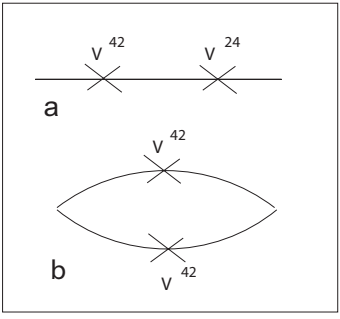}}
\caption{ Diagrams for the second iteration of the perturbation theory; corrections to the Green's function (a), corrections to the vertex in  conductivity (b).}
\label{diag}\end{figure}  
 
 A simplest way to evaluate the corrections  resulting from the warping $\gamma_3$ consists in the consideration of the relative Green's function having the poles at the electron levels.
The corrections to the levels can be found in the iterations  
 \begin{equation}{\bf G}_{m+1}(x,x')=\int dx''{\bf G}_0(x,x''){\bf V}(x''){\bf G}_m(x'',x')\,,\label{it}\end{equation}
where $\bf V(x)$ has only two matrix elements $V^{42}=\omega_c\tilde{\gamma_3}a^{+}/\gamma_0$ and $V^{24}=V^{42*}$ in the  Hamiltonian (\ref{ham}).
  The Green's function of the unperturbed  Hamiltonian  writes 
 \begin{equation}
 G^{\alpha \beta}_0(\varepsilon,x,x')=\sum_{sn}\frac{\psi^{\alpha}_{sn}(x) \psi^{*\beta }_{sn}(x')}{\varepsilon-\varepsilon_{sn}}\,.\label{gf}\end{equation}

In the second iteration, we 
get the corrections
\[\int dx_1 dx_2G_0^{\alpha4}(x,x_1)V^{42}(x_1)G_0^{22}(x_1,x_2)V^{24}(x_2)G_0^{4\beta}(x_2,x')\]
and the similar term with the substitution of the superscripts $2\leftrightarrow4$. The matrix elements of the perturbation $V$ are easily calculated with respect to the Hermitian functions of Eqs. (\ref{gf}), (\ref{func}) and we obtain for the diagram shown in the upper part of Fig. \ref{diag} 
\begin{equation}\left(\frac{\omega_c\tilde{\gamma}_3}{\gamma_0}\right)^2\sum_{s'sn}\frac{(n-2)|C^4_{sn}C^2_{s',n-3}|^2\psi^{\alpha}_{sn}(x) \psi^{*\beta }_{sn}(x')}{(\varepsilon-\varepsilon_{sn})(\varepsilon-\varepsilon_{s',n-3})(\varepsilon-\varepsilon_{s,n})}.\label{corr1}\end{equation}
This correction plays an important role near the poles of the Green's
function. Therefore, we can substitute $\varepsilon_{sn}$ instead of $\varepsilon$ in the second factor of the denominator and represent this correction as a shift $\delta\varepsilon_{sn}$ of the poles $(\varepsilon-\varepsilon_{sn}-\delta\varepsilon_{sn})^{-1}$ with
 \begin{equation}
 \begin{array}{c}
{\displaystyle\delta\varepsilon_s(n)=\left(\frac{\omega_c\tilde{\gamma}_3}{\gamma_0}\right)^2\sum\limits_{s'}\left\{\frac{(n-2)|C^4_{sn}C^2_{s',n-3}|^2}{\varepsilon_{s}(n)-\varepsilon_{s'}(n-3)}\right.}\\ 
+{\displaystyle\left.\frac{(n+1)|C^2_{sn}C^4_{s',n+3}|^2}{\varepsilon_{s}(n)-\varepsilon_{s'}(n+3)}\right\} }\,,
\end{array}
\label{pt1}\end{equation}
where the first term should be omitted for $n-3<0$. 
In fact, our illustration is nothing but a calculation of the electron self-energy and the naive expansion of the denominator can be indeed replaced by  summarizing of the corresponding diagrams.


Comparing
the corrections, Eq. (\ref{pt1}), with the main contribution Eq. (\ref{de1}), we find that the
perturbation theory is valid  when an  expansion parameter $(\tilde{\gamma}_3\tilde{\gamma}_1/\gamma_0\omega_c)^2$ becomes small, i.e. for the strong magnetic fields $B> 1 T$.
  The corrected $|10\rangle$ level  writes \begin{equation}\varepsilon_1(n=0)=\tilde{\gamma}_2+\left(\frac{\omega_c\tilde{\gamma}_3}{\gamma_0}\right)^2\sum\limits_{s'}\frac{|C^4_{s'3}|^2}
{\tilde{\gamma}_2-\varepsilon_{s'}(3)}\,.\label{nc0}\end{equation}   
  The $|21\rangle-$level 
  is very close to the level with $n=0$, Eqs. (\ref{n0})  and (\ref{nc0}). 
  
  Our expressions for the levels with the corrections (\ref{pt1}) and (\ref{nc0}) give the same results as obtained in Ref.
  \cite{Na} by the numerical method of truncating the infinite matrix.


 \section{conductivities in magnetic fields}  
 
 In the collissionless limit  when the relaxation rate $\Gamma$ is much less than the frequency, $\Gamma\ll\omega$, the conductivity is expressed in terms of the correlation function 
\begin{equation}
\mathcal{P}\left( \omega
\right) =T\sum\limits_{\omega
_{m}}\int dx dx'Tr\left\{ v^{i}%
G\left( \omega_{+},x,x'\right) v^{j}G\left( \omega_{-},x',x\right)
\right\} \label{cor} \end{equation}
 where we should (i) summarize over Matsubara's frequencies $\omega_m$, (ii) take the trace over
 the Landau and band numbers, (iii) make an analytic continuation into real frequencies $\omega$,
 and (iiii) substrate from the result its value at $\omega=0$ (for details see Ref. \cite{AGD,FV}).

The velocity matrices  $v^i$ in Eg. (\ref{cor}) are given by the derivative of the  Hamiltonian, Eq.~ (\ref{ham}),
\begin{equation}
\mathbf{v}=\frac{\partial H(\mathbf{k})}{\partial \mathbf{k}}\,.
\label{vel}
\end{equation}%

First we consider  the largest velocity operators, Eq. (\ref{vel}), which do not involve  the parameter $\tilde{\gamma}_3/\gamma_0$\,.     Straightforward calculations yield two independent components of the dynamical conductivity 
\begin{equation}
\begin{array}{c}
\left.\begin{array}{c} \sigma_{xx}(\omega)\nonumber\\ i\sigma_{xy}(\omega)
\end{array}\right\}=i{\displaystyle\sigma_0
\frac{4\omega_c^2}{\pi^2}}
{\displaystyle\sum_{n,s,s'}\int\limits_0\limits^{\pi/2}dz\frac{\Delta f_{ss'n}}{\Delta{ss'n}}|d_{ss'n}|^2}\\ 
\times
\left[(\omega+i\Gamma
+\Delta_{ss'n})^{-1}\pm
(\omega+i\Gamma-\Delta_{ss'n})^{-1} \right]
\,,
\end{array}
\label{dc1}\end{equation}
where  
$\Delta_{ss'n}=\varepsilon_{sn}-\varepsilon_{s', n+1}$ is the level spacing including the corrections given in Eqs. (\ref{pt1}) and (\ref{nc0}), $\Delta f_{ss'n}=f(\varepsilon_{s'n+1})-f(\varepsilon_{sn})$ is the difference of the corresponding Fermi-Dirac functions and 
\begin{equation}\begin{array}{c}
d_{ss'n}=C^2_{sn}C^{1}_{s'n+1}+C^{3}_{sn}C^{4}_{s'n+1}\nonumber\\
+(\tilde{\gamma}_4/\gamma_0)(C^1_{sn}C^{4}_{s'n+1}+C^{2}_{sn}C^{3}_{s'n+1})\end{array}\label{dip}\end{equation} is the
dipole matrix element.  These electron transitions obey the selection rule
$$\Delta n=1\,,$$
and  will be referenced as the strong lines. The integration over the Brillouin half-zone, $0<z<\pi/2$,  and the summation over the Landau number $n$ as well as  the bands $s, s'$ should be done in Eq. (\ref{dc1}). 
The conductivity units  
$$\sigma_0=\frac{e^2}{4\hbar d_0}$$
have a simple meaning, being  the  graphene dynamic conductivity \cite{Kuz}   $e^2/4\hbar$ multiplied by the number  $1/d_0$ of layers within the  distance unit in the main axis-direction.  

Now we consider the terms with $\tilde{\gamma}_3/\gamma_0$\, in the velocity operators, Eq. (\ref{vel}). Calculating the correlation function Eq. (\ref{cor}) we get an additional term in the conductivity, which can be obtained from Eq. (\ref{dc1}) with the substitutions $$n+1\rightarrow n+2$$   and with the matrix element  
$$d_{ss'n}= (\tilde{\gamma}_3/\gamma_0)C^2_{sn}C^{4*}_{s'n+2}$$
instead of the matrix element given by Eq. (\ref{dip}). This transitions obey   the selection rule $$\Delta n=2$$ 
and will be referenced as the weak lines.

So far we considered the $\gamma_3$ corrections to the Green's function, i.e. to the levels.  However, there are so-called vertex corrections  to the self-energy shown at the bottom of Fig. \ref{diag}. They are resulted from the  quartet of the coupled Landau levels, which interfere while  the selection rules $\Delta n=1$ and $\Delta n=2$ are allowed.
For compactness, let us denote  this quartet of given $n$   as following
 \begin{equation} 
a=|sn\rangle,\, b=|s',n+1\rangle,\, c=|s_1,n+3\rangle,\,
 d=|s_1',n+4\rangle
 \,,
 \label{lev} \end{equation}
   where the band numbers $s, s', s_1$,  and $s_1'$ are arbitrary. 

The corresponding corrections to  conductivities write
\begin{displaymath}\begin{array}{c}
\left.\begin{array}{c}\delta \sigma_{xx}(\omega)\nonumber\\ i \delta\sigma_{xy}(\omega)
\end{array}\right\}=2i\sigma_0{\displaystyle
\sum_{nss's_1s_1'}\int\limits_0\limits^{\pi/2}dz \left(\frac{2\omega_c^2\tilde{\gamma}_3}{\pi\gamma_0}\right)^2}
\\{\displaystyle
\times 
C^2_aC^2_bC^4_cC^4_d(C^1_bC^2_a+C^3_aC^4_b)(C^1_dC^2_c+C^3_cC^4_d)}
\\ \times \sqrt{(n+1)(n+2)}
(\varepsilon_b-\varepsilon_d)^{-1}(\varepsilon_a-\varepsilon_c)^{-1}\\
\times\left\{\left[(\omega+i\Gamma+\varepsilon_b-\varepsilon_a)^{-1}
\pm(\omega+i\Gamma-\varepsilon_b+\varepsilon_a)^{-1}\right]\partial_{ab}\right.
\\+\left[(\omega+i\Gamma+\varepsilon_b-\varepsilon_c)^{-1} \pm(\omega+i\Gamma-\varepsilon_b+\varepsilon_c)^{-1}\right]\partial_{cb}\\+
\left[(\omega+i\Gamma+\varepsilon_d-\varepsilon_a)^{-1}
\pm(\omega+i\Gamma-\varepsilon_d+\varepsilon_a)^{-1}\right]\partial_{ad}
\\ \left. \nonumber+\left[(\omega+i\Gamma+\varepsilon_d-\varepsilon_c)^{-1} \pm(\omega+i\Gamma-\varepsilon_d+\varepsilon_c)^{-1}\right]\partial_{cd}\right\}
\end{array}\end{displaymath}
where 
$$\partial_{ab}=[f(\varepsilon_a)-f(\varepsilon_b)]/(\varepsilon_b-\varepsilon_a)$$ and $f(\varepsilon_a)$ is the Fermi-Dirac function.
The terms with the negative radicand should be omitted while summing over $n$ and all band numbers
from Eq. (\ref{lev}).  
\begin{figure}[]
\resizebox{.52\textwidth}{!}{\includegraphics{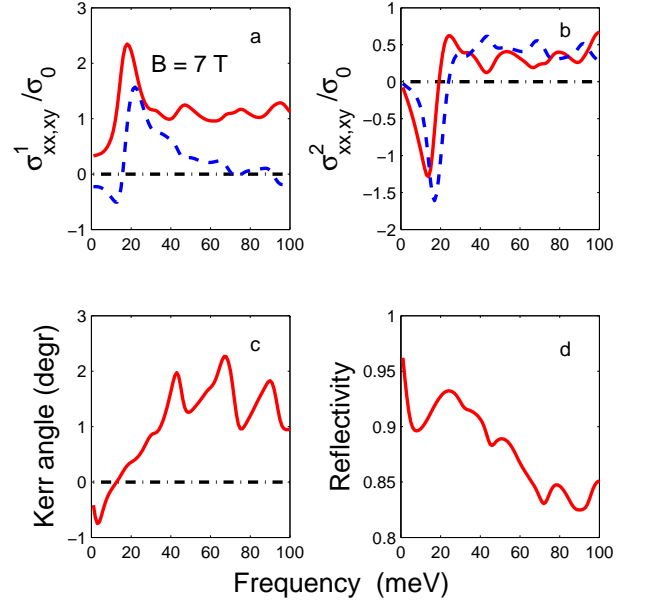}}
\caption{(Color online)  Real (a) and  imaginary (b) parts of the longitudinal (xx, solid line)  and Hall (xy, dashed line) dynamical conductivities; Kerr angle (c) and reflectivity (d).
The magnetic field  $B=$ 7 T, the temperature T = 0.1 meV is less than the level broadening $\Gamma=3.5$ meV.}
\label{xx7}
\end{figure}
\section{Hall and longitudinal conductivities with the SWMC parameters}

The parameters of Eq. (\ref{ham}) used in the calculations are listed in Tabl. \ref{tb1} (see also Ref. \cite{BCP}). The  hopping integrals $\gamma_0$ to
$\gamma_3$ are close to the values determined in observations of the semiclassical ShdH effect. The Fermi energy equal to $\varepsilon_F=-4.1$ meV  agrees at the zero magnetic field   with the measurements of the extremal Fermi-surface cross sections and the masses of holes and electrons. 
Connections with the notation for the same parameters in the SWMC model are given in the "SWMC" line.
  The values of parameters $\gamma_4$, $\gamma_5$, and $\Delta$ determined in various experiments are very different, we use $\gamma_5$ and $\Delta$  obtained by Doezema et al \cite{DDS} (given in Tabl. \ref{tb1} in the "SWMC" notations) and take for $\gamma_4$ the approaching value. In the quantum limit, when electrons and holes occupy only  $|10\rangle$ and $|21\rangle$ levels, the Fermi energy must cross these close levels at the middle of the KH line. It means that the Fermi level becomes higher at such the magnetic fields   taking the value $\varepsilon _F\approx -1$ meV.
\begin{figure}[]
\resizebox{.5\textwidth}{!}{\includegraphics{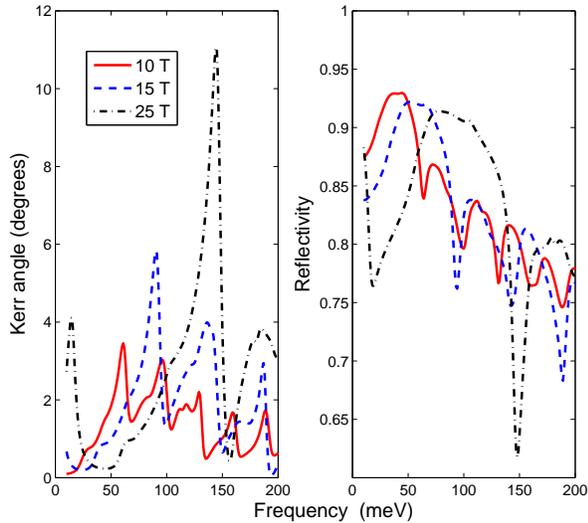}}
\caption{(Color online) Kerr angle and reflectivity at  10, 15, and 25 T.}
\label{kerrang1}
\end{figure} 

The results of calculations are represented in Figs. \ref{xx7}-\ref{kerrang1}. Let us emphasize that
 the imaginary part of the dynamical conductivity is of the order of the real part.

One can see in Fig.  \ref{xx7} (a), that  the longitudinal conductivity calculated per one
graphite layer tends on average  to the  graphene universal conductance.
The main contribution in the sharp 16-meV line is resulted from
the electron $|21\rangle\rightarrow|32\rangle $ transitions (15 meV) about the $K$ point  where the $|32\rangle $ level coincides with the Fermi level (within  accuracy of the width $\Gamma$ or temperature $T$). 
 Then, the transitions $|22\rangle\rightarrow|21\rangle $ produce 
the broad band. The low-frequency side of the band (23 meV, at the intersection of the  $|21\rangle $ level with the  Fermi level) contributes into  the  16-meV line. 
In the same 16-meV line, the transitions $|32\rangle\rightarrow|33\rangle $ can contribute as well if the band $|32\rangle$ contains the electrons.

The next doublet at 43 meV  arises  from the transitions $|23\rangle\rightarrow|32\rangle $  and $|22\rangle\rightarrow|33\rangle$  at the $K$ point.
The 68- meV doublet splitting results due to the electron-hole asymmetry   from the transitions $|24\rangle\rightarrow|33\rangle $ (65 meV) and $|23\rangle\rightarrow|34\rangle$ (69 meV) at the $K$ point  of the Brillouin zone.   
 
 The 89-meV line is more complicated. First,  there are the electron transitions  $|24\rangle\rightarrow|35\rangle\,$ (89 veV) and $|25\rangle\rightarrow|34\rangle $ (90 meV) near the $K$ point. 
  Besides,   the transitions $|11\rangle\rightarrow|10\rangle $ (95 meV) near the $H$ point make  a contribution as well. All these lines  obeying the selection rule $\Delta n=1$ are strong. There are two weak lines in the frequency range. One ($|24\rangle\rightarrow|32\rangle $) is seen at 55 meV 
  as a shoulder  on the theoretical curve. Another, at 31 meV, results from the transitions $|10\rangle\rightarrow|32\rangle $ near the $K$ point.
 
The positions of the lines  for  fields in the range of 10 -- 30~ T agree   with observations of Refs. \cite{OFS,CBN}.
   
The optical Hall conductivity $\sigma_{xy}(\omega)$ in the ac regime is shown in Fig.   \ref{xx7} (a) and (b). 
The conductivities 
$\sigma_{xx}(\omega)$ and $\sigma_{xy}(\omega)$ allow us to calculate the Kerr rotation and the reflectivity  as functions of frequency [see  Fig. \ref{xx7} (c) and (d)]. 
It is evident that the interpretation of the Kerr rotation governed by
the conductivity $\sigma_{xy}(\omega)$  is much more complicated
  in comparison with the longitudinal conductivity.  
 
 The Kerr angle and reflectivity shown in Fig. \ref{kerrang1} for the different magnetic fields demonstrate the strong field dependence of the magneto-optic phenomena.

\section{summary and conclusions}

In conclusions, we have evaluated the perturbation theory for the matrix Hamiltonian, which permits to calculate the corrections to the eigenvalues resulting from the small matrix elements particularly  from the trigonal warping. We have shown that the trigonal warping in graphite can be considered within the perturbation theory at the strong magnetic field larger than 1 T approximately. We have found that the main optical transitions obey the selection rule $\Delta n=1$ for the Landau number $n$, however the $\Delta n=2$ transitions due to the trigonal warping with the small probability are also essential. The good agreement between the calculations and the measured  Kerr rotation and reflectivity in graphite in the quantizing magnetic fields is achieved. The SWMC model parameters are used in the fit  taking into account the previous dHvA measurements  and correcting the Fermi energy for the case of the strong magnetic fields. 

\acknowledgments
The author  is thankful to A. Kuzmenko and J. Levallois for useful discussions 
and for communicating their experimental results 
prior to publication.
This work was supported by the Russian Foundation for Basic
Research (grant No. 10-02-00193-a) and by the SCOPES grant IZ73Z0$\_$128026 of the Swiss NSF. The
author is grateful to the Max Planck Institute for  Physics of
Complex Systems for its hospitality in Dresden.

 \end{document}